# Van der Waals heterojunction devices based on organohalide perovskites and two-dimensional materials

*Hung-Chieh Cheng,[†] Gongming Wang,[‡] Dehui Li,[‡] Qiyuan He,[‡] Anxiang Yin,[‡] Yuan Liu,[†] Hao Wu,[†] Mengning Ding,[†] Yu Huang,[†,§] and Xiangfeng Duan*[‡,§]*

[†]Department of Materials Science and Engineering, [‡]Department of Chemistry and Biochemistry, University of California, Los Angeles, CA 90095, USA

[§]California Nanosystems Institute, University of California, Los Angeles, CA 90095, USA

The recently emerged organohalide perovskites (e.g., $CH_3NH_3PbI_3$) have drawn intense attention for high efficiency solar cells. However, with a considerable solubility in many solvents, these perovskites are not typically compatible with conventional lithography processes for more complicated device fabrications that are important for both fundamental studies and technological applications. Here we report the creation of novel heterojunction devices based on perovskites and two-dimensional (2D) crystals by taking advantage of the layered characteristic of lead iodide ($PbI_2$) and vapor phase intercalation. We show a graphene/perovskite/graphene vertical stack can deliver a highest photoresponsivity of ~950 A/W and photoconductive gain of ~2200, and a graphene/$WSe_2$/perovskite/graphene heterojunction can display a high on/off ratio (~$10^6$) transistor behavior with distinct gate-tunable diode characteristics and open-circuit voltages. Such unique perovskite-2D heterostructures have significant potential for future **optoelectronic research and can enable broad possibilities with compositional tunability of organohalide perovskites and the versatility offered by diverse 2D materials.**

The recently discovered organic-inorganic hybrid perovskites (*e.g.*, $CH_3NH_3PbI_3$) have been demonstrated to be an attractive material for solution-processed solar cell[1-15] with a record high power conversion efficiency (PCE) of ~20.1%.[11, 13] Such high PCE have been generally attributed to their strong light absorption[16] and long free carrier diffusion length.[3, 9, 10, 17] In addition to solar cell, the perovskites have also been explored for several applications, such as light-emitting diodes,[18, 19] lasers[20, 21] and photodetectors,[12, 22-24] which have rendered them promising candidates in optoelectronic applications. However, the solubility of $CH_3NH_3PbI_3$ in many common solvents including water, acetone and alcohol, and their instability in ambient condition have impeded their development in conventional device fabrication, and the studies to date have largely been limited to spin-coated perovskite thin films.[19, 22, 25] To fully explore their electronic and optoelectronic properties, it is important to integrate the perovskite materials into different device structures. Here we propose a new approach for the fabrication of complex perovskite devices, and demonstrate a new type of vertical photoconductive and photovoltaic perovskite devices with graphene as electrodes through a layer-by-layer *van der Waals* assembly process.[26-29] Taking advantage of the layered characteristics of lead iodide ($PbI_2$),[30, 31] we are able to exfoliate $PbI_2$ crystal (Fig. S1) into thin flakes (Fig. S2) that can then be converted into $CH_3NH_3PbI_3$ under methylammonium iodide (MAI) vapor phase reaction.[2] Fig. 1a, b shows the comparison of mechanically-exfoliated $PbI_2$ flakes before and after the conversion process. With the intercalation of MAI in $PbI_2$, the averaged flake thickness increased around 1.9 times from 22.7 nm to 43.2 nm (Fig. 1c), which is comparable to previous report on synthetic nanoplatelets.[32] Such transformation accompanied with increased thickness can be attributed to the volume expansion results from organic molecule intercalation (Fig. 1d). Moreover, Fig. 1e shows the photoluminescence (PL) spectrum taken from the same spot before and after conversion. A prominent single PL peak around ~1.63 eV appeared after conversion process, which indicates a successful conversion from $PbI_2$ to $CH_3NH_3PbI_3$.

In order to understand the basic photoconductive properties of $CH_3NH_3PbI_3$, we have first created a graphene/perovskite/graphene (GPG) heterostructure device, in which graphene is used as both the top- and bottom-electrodes to sandwich the perovskite in the middle. Instead of using the perovskite as the sensitizer[23] and measuring the conductivity change in graphene, we measured the current transport from the top- to bottom-graphene directly through the perovskite channel. Since both $PbI_2$ and perovskite are quite sensitive to the moisture in ambient condition, and soluble in common solvents that are usually used in fabrication process. We adopted a particular dry-process for creating designed heterostructures as described in supporting information (Fig. S3). Fig. 2a shows schematic illustration and an optical image of an as-fabricated GPG device. Electrical characterization was conducted in vacuum at room temperature. Fig. 2b shows the current-voltage ($I_{DS}$-$V_{DS}$) characteristics measured under 532 nm laser with an equivalent power ~1.52 nW on the device area. By applying a constant bias voltage, the device worked as a photoconductor, the photocurrent $I_{ph} = (I_{illuminated} - I_{dark})$ increased with bias voltage as a consequence of the reduction of the carrier transit time. The photocurrent at bias 1 V is ~1.45 µA, which yields a photoresponsivity ~950 A/W and a photoconductive gain ~2200. However, during the time-domain measurement, the current drop quickly before reaching a stable state under a constant bias, which could be attributed to the ion movement in perovskite material driven by the electric field.[22] Fig. 2c shows the photocurrent response versus time after the current reaches a stable state. The photocurrent rise time is ~22 ms and the fall time is ~37 ms, which is reasonable for a high gain photoconductor. In order to spatially characterize the photocurrent contribution, we performed the spatially-resolved photocurrent map under focused laser beam (488 nm) in air (Fig. 2d). The photocurrent map results show that the photocurrent is uniformly contributed from the graphene/perovskite/graphene overlapping area, indicating that the majority of photo-induced carriers travel in vertical direction.

Despite the protection by the top-graphene, the GPG devices still suffer from the attack of moisture during the measurement in air. To further improve device stability, we have transferred another boron nitride (BN) layer as an ultrathin protection layer for the GPG stack. The protection by BN can also allow us using subsequent lithography process for the fabrication of GPG device. Since the top- and bottom-graphene may be covered by the BN, we have used edge metal contacts[26, 27] for both the top- and bottom-graphene electrodes (see Methods). Fig. 3a shows the schematic illustration and optical image of BN-covered GPG device. Our study has proved that BN-protected devices show no considerable degradation in terms of photocurrent characteristics after 210 days from the first measurement (Fig. S4), demonstrating BN can serve as an effective protection for organohalide perovskite to significantly extend the lifetime of GPG device.

To explore semiconducting characteristics of perovskite, we measured the device current under different back-gate voltage. However, there is no obvious gate-modulation observed in our device at room temperature, which could be attributed to the gate-induced ion movement in $CH_3NH_3PbI_3$ compensating/screening the electric field from the back gate.[19] To minimize the effect of ion movement, we have conducted the measurement at 77 K. The $I_{DS}$-$V_{DS}$ characteristics under dark show asymmetric response to the gate voltage (Fig. 3b), which can also be identified in the transfer curves (Fig. 3c). It is evident that the charge transport across the vertical stack increases with increasing negative gate voltage, suggesting a p-type hole transport behavior. Overall, there is considerable gate response when a negative bias is applied to the top-graphene electrode, while there is little gate-modulation when the top-graphene electrode is positively biased. The transfer curves obtained in the dark show that the current on/off ratio is only about ~5 when the top-graphene electrode is positively biased (Fig. 3c, red curve), and can reach up to ~500 while the top-graphene electrode is negatively biased (Fig. 3c, black curve). The reasons for such asymmetric effect can be explained by the gate-modulation to the graphene work function and the Schottky barrier for hole-injection at bottom-graphene/perovskite interface. With negative bias applied to the top-graphene electrode, the holes are injected from bottom-graphene to perovskite and thus the bottom-graphene/perovskite barrier dominates the charge transport. In this case, a negative (positive) gate voltage elevates (reduces) the graphene work function and decreases (increases) the Schottky barrier for hole injection from the bottom-graphene to the perovskite, and thus enhancing (suppressing) the current across the vertical stack (depicted in Fig. 3c inset). On the other hand, with a positive bias applied to the top-graphene, the holes are injected from top-graphene to the perovskite, where the barrier is only weakly modulated by the back-gate bias, and therefore we expect much smaller gate-modulation of the current transport. These analyses also further confirm that the holes are the dominated carrier in our GPG devices.

    We have also investigated the photovoltaic effect in our GPG device. The $I_{DS}$-$V_{DS}$ curves are measured at varying gate voltages under laser illumination (Fig. 3d), and plotted in the

logarithmic scale to examine the open-circuit voltage ($V_{OC}$). Importantly, as the gate voltage is increased from -60 V to 60 V, we observed a consistent shift of $V_{OC}$ from a negative value to a positive value (Fig. 3e). To exclude the possibility of device instability or time-dependent $V_{OC}$ shift, we performed the same measurement with reverse gate voltage sequence from 60 V to -60 V (Fig. S5). Our studies show that $V_{OC}$ shift is consistent with the applied gate bias rather than sequence in the measurement, indicating the shift is indeed resulted from the gate-modulation. Such gate-modulation effect can be attributed to the asymmetric tuning of top- and bottom-graphene work function, which induced a built-in electric field across the entire GPG structure. Basically, an applied back-gate voltage can electrostatically dope bottom-graphene and therefore modify its work function, and thus creating a work-function difference between top- and bottom-graphene, which resulted in a built-in electrical field that can drive the separation and transport of the photogenerated electrons and holes in opposite direction to produce a non-zero photogenerated, short-circuit current ($I_{SC}$) at zero-bias condition. As depicted in Fig. 3e inset, when $V_G < 0$ ($> 0$), an additional negative (positive) bias to top-graphene is needed to compensate the built-in field leading to negative (positive) $V_{OC}$. In the context of GPG symmetric vertical structure, the $V_{OC}$ and its gate-tunability are considerably smaller than other system such as graphene/MoS$_2$/graphene,[33] which may be attributed to its distinct ion movement effect under applied gate-voltage. Considering the case when the negative (positive) gate voltage is applied (Fig. 3e), the positively (negatively)-charged ions in perovskite tend to accumulate at the interface between bottom-graphene and perovskite, which counters the gate-field induced work function modulation in bottom-graphene and results in less built-in field and smaller $I_{SC}$ in this GPG device.

To further increase the built-in potential across the GPG device and explore more tunability with gate voltage, we have inserted an extra WSe$_2$ layer between bottom-graphene and perovskite to create a graphene/WSe$_2$/perovskite/graphene (GWPG) device (Fig. 4a). By applying bias to the top-graphene with bottom-graphene grounded, we have investigated the transfer characteristic of this GWPG vertical device as shown in Fig. 4b. We found the on/off ratio of negatively-biased device has been greatly improved to ~10$^6$ compared to that in the previous symmetric GPG device (~500) measured in the dark. Such distinct gate-tuning effect can be considered as the combination of the gate-modulation of the Schottky barrier height at bottom-graphene/WSe$_2$ (GW) interface and the modulation of the WSe$_2$/perovskite (WP) heterojunction interface. To elucidate the gate-modulation effect, we are now focusing on the $I_{DS}$-$V_{DS}$ characteristics under dark condition (Fig. 4c). A nearly symmetric I-V curve was obtained with gate voltage within the range from -60 V to -20 V, while gradually evolved into rectifying diode behavior as the gate voltage shifts toward more positive direction. Such effect can be attributed to the transition from p-p to n-p junction at WP interface due to the ambipolar nature of WSe$_2$ under different gate bias.[34] Indeed, our control experiment on graphene/WSe$_2$/graphene (GWG) sandwich device confirms that similar ambipolarity can be observed in WSe$_2$ in the vertical configuration (Fig. S6). In this way, under a large negative gate bias, the WSe$_2$ is p-type and therefore forms a p-p junction with the p-type perovskite. With increasing gate voltage toward positive direction, the WSe$_2$ is turned into n-type material, and therefore forms an n-p junction with the p-type perovskite.

In Fig. 4d and 4e, we consider the device under illumination. While the heterojunction device is within the p-p region ($V_G$ = -60 V to -20 V), a small $V_{OC}$ and weak gate-modulation were observed, which can be explained by the relative small built-in potential across WP junction since both WSe$_2$ and perovskite are p-doped by the gate bias. With the $V_G$ applied from -20 V to 0 V, a dramatic change occurs due to the switch of carrier type in WSe$_2$. The WSe$_2$ is now becoming n-type and forming n-p junction with p-type perovskite. As the positive $V_G$ increases, the WSe$_2$ is more n-doped, while the perovskite remains p-type, which further increases the $V_{OC}$. However, the increasing positive $V_G$ also reduces the p-type nature of perovskite, thus a smaller gate tuning $V_{OC}$ slope was observed in n-p region compared to the transition region (Fig. 4e). It is interesting to note the $V_{OC}$ shows most dramatic change while the GWPG device is being switched from the p-p junction to n-p junction. To further look into the gate modulation of $V_{OC}$, we have plotted the absolute current of the device versus gate voltage and source-drain bias in a two-dimensional color-contour plot (Fig. 4f). The current minimum (corresponding to $V_{OC}$) shows an obvious dependence on the gate voltage, highlighting the gate modulation of $V_{OC}$. The same gate-dependent measurement under dark condition shows no drift of current minimum with $V_G$, (Fig. 4f inset), further confirming the drift of current minimum under light illumination is resulted from the gate-tuning of the photo-induced $V_{OC}$. It should be noted that such p-p to n-p transition was not observed in room temperature measurement. At room temperature, the ion motion mediated by vacancies[35] in perovskite is able to compensate the gate-field and thus prevent gate-modulation of WSe$_2$ from p-type to n-type. This has also verified the ion motion plays a substantial role in perovskite-based device. We believe more fundamental studies and continued optimization of these unique perovskite-2D heterostructure are needed for designing high performance optoelectronic applications.

In summary, we have developed a unique approach to integrate perovskite with 2D materials. We realized a high photoconductive gain up to ~2200 and a high photoresponsivity of ~950 A/W in a vertical GPG device structure. With the effective protection of BN for perovskite, we are able to extend the lifetime of such perovskite-2D material heterostructures up to 7 months. The low temperature measurement revealed the possibility of gate-tunable photovoltaic behavior based on perovskite devices, and offered important insights on how the ion motion affect the electrical properties of GPG based vertical device. The GWPG devices uncovered the potential to build a high on/off ratio vertical transistor based on perovskite materials and a gate-tunable photodiode by integrating other 2D materials beyond graphene. By combining their advantages in either optoelectronic or electronic properties, we are able to explore more underlying physical properties of this new promising perovskite material. In the future, we believe such platform is interesting and versatile. In additional to integrate with various 2D materials, the organohalide perovskites themselves have great tunability in band gap by engineering their composition.[5] One feasible way is to replace MAI with other halides or different halogenated organic molecules to react with exfoliated starting material $PbI_2$. Further investigation on such perovskite-2D material heterostructures will be beneficial for designing next generation hybrid optoelectronic device.

**Methods.**

*Device fabrication.* The $PbI_2$ crystals for mechanical exfoliation are prepared by re-crystallization of commercial available $PbI_2$ powder. 0.4 g $PbI_2$ powder is dissolved in 100 mL water at 100 °C. While resulting $PbI_2$ hot solution is naturally cooled to room temperature, the $PbI_2$ crystals nucleate and grow into hexagonal crystals (Fig. S1). The solution with precipitated $PbI_2$ crystals is then dried in oven for 30 minutes. The completely dried crystals as shown in (Fig. S1) are now ready for exfoliation and stack of van der Waals heterostructures. A layer of PMMA is first spun on a silicon substrate, followed by coating another layer of poly-propylene carbonate (PPC). The PbI$_2$ typically with a thickness of 50-80 nm are exfoliated onto the PMMA/PPC polymer stack. Then polymer stack is peeled off from silicon substrate and attached to a PDMS (polydimethylsiloxane) stamp with PbI$_2$ side up. The PDMS stamp with PbI$_2$ sitting on PMMA/PPC stack is positioned on a prepared monolayer graphene exfoliated on 290 nm SiO$_2$/Si substrate. Then PDMS and the PbI$_2$ are brought to contact with graphene. The polymer stack is then released from PDMS stamp by heating up the substrate to 120°C. After transfer, the polymer stack is dissolved in chloroform for 10 minutes. The PbI$_2$ sitting on graphene is then converted into perovskite using vapor phase reaction with methylammonium iodide vapor carried by argon gas flow. Another graphene serving as top electrode is prepared on another PMMA covered silicon substrate, which is transferred immediately on the as-converted perovskite by the same method described above. The carrier PMMA itself can serve as the resist for e-beam lithography. The device is then prepared after a vacuum metallization (Cr/Au: 20 nm/80 nm) and lift-off process in chloroform. For BN-covered GPG device, the carrier PMMA is dissolved in chloroform after top-graphene transfer. Subsequently, another separately prepared boron nitride (BN) is used to cover the entire area of perovskite. Sometimes the graphene electrodes are fully covered by BN. To make electrical contacts with BN covered graphene, we create edge contacts by etching the BN with fluoroform/O$_2$ (35/5 sccm) plasma for 1 min. The BN etching rate is faster than graphene, so we can always make good electrical contacts with graphene. For graphene/WSe$_2$/perovskite/graphene (GWPG) device, the WSe$_2$ (~50 nm) flakes are first transferred onto exfoliated monolayer graphene using PMMA/PPC stack. After dissolve polymer stack in chloroform, the WSe$_2$/graphene stack is now ready for PbI$_2$ transfer. The following steps are same as those described above.

*Device characterization.* The electrical and low temperature measurements were both carried out in commercial probe station (Lakeshore, TTP4) equipped with a 532 nm laser (Coherent, 532-100). The electrical measurement was conducted with precision source/measure unit (Agilent, B2902A). The photoluminescence spectra were taken with a commercial confocal micro-Raman system (Horiba LABHR) excited by the Ar-ion laser (488 nm) with an excitation power ~3.8 µW. The photocurrent map was conducted in the same system by monitoring the current under the irradiation of focused laser (488 nm, < 300 nW) while moving the stage within a specified area. The AFM images were taken from Bruker Dimension 5000 Icon scanning probe microscope.

**Figures**

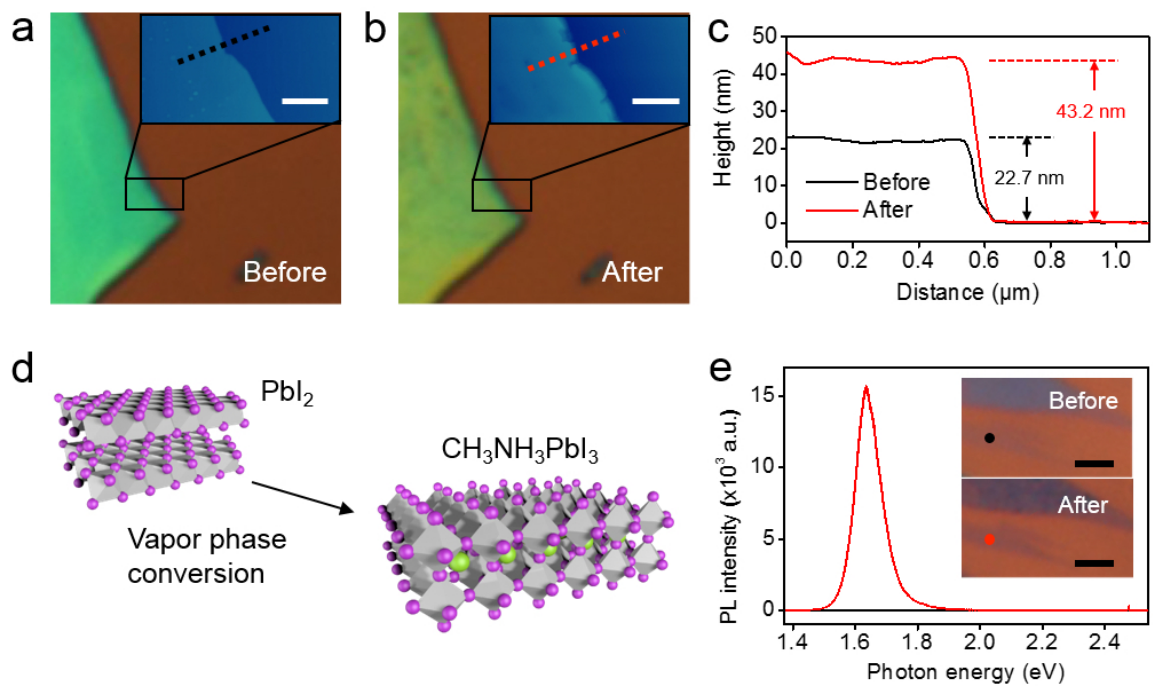

**Figure 1.** Conversion of exfoliated lead iodide flake into perovskite. (a, b) The optical images of an exfoliated lead iodide flake before and after conversion process, the corresponding insets show the AFM images of the area marked by black rectangles in optical images. The white scale bars for AFM images in (a) and (b) are both 500 nm. (c) The sectional height profile extract from black and red dashed lines from the insets of (a) and (b), respectively. An increased thickness was observed after conversion. (d) Schematic of the structure change from layered $PbI_2$ to perovskite. (e) The black and red curves represent the corresponding photoluminescence spectra taken from the black and red spots shown in insets (optical images), respectively. A prominent emission peak at 1.63 eV was detected after conversion, which indicates a successful perovskite formation.

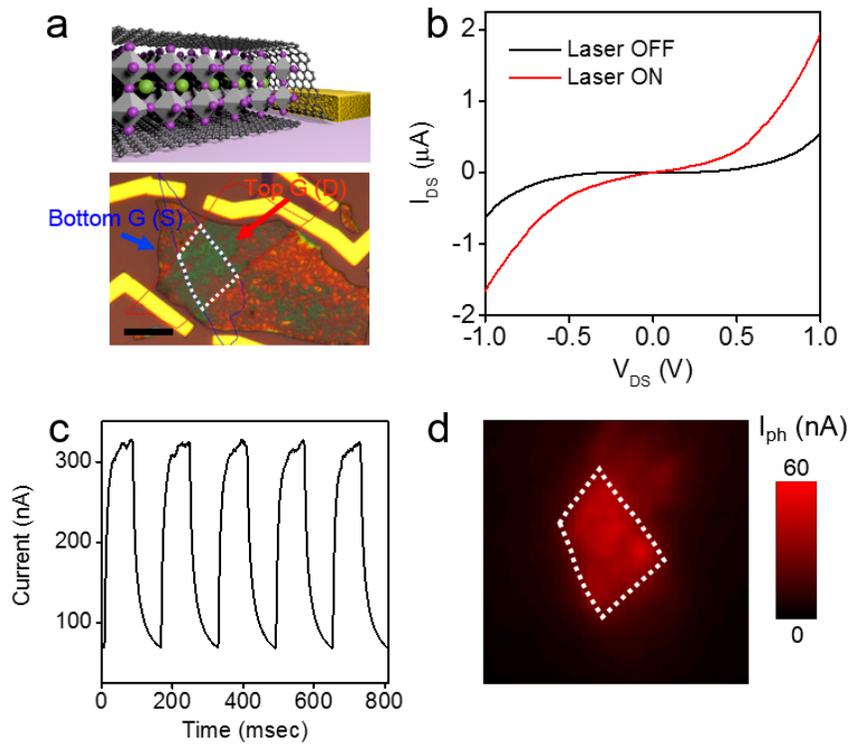

**Figure 2.** Photoconductive characterization of graphene/perovskite/graphene (GPG) vertical device. (a) Schematic illustration and optical image of a GPG heterostructure device. The white dashed line outlines the overlapping area, the bottom-graphene and top-graphene are outlined by the blue and red solid lines, respectively. The top-graphene is biased (drain) while the bottom graphene is grounded (source) during all measurement. The black scale bar is 10 μm. (b) $I_{DS}$-$V_{DS}$ curves of GPG device measured with/without laser illumination. A photo-induced current offset was observed. (c) The time-domain photocurrent response measured with constantly applied bias while illuminated by 532 nm laser (chopping frequency at 7 Hz). (d) The corresponding photocurrent map for the device shown in (a). The major contribution of photocurrent was from the outlined GPG overlapping area in (a).

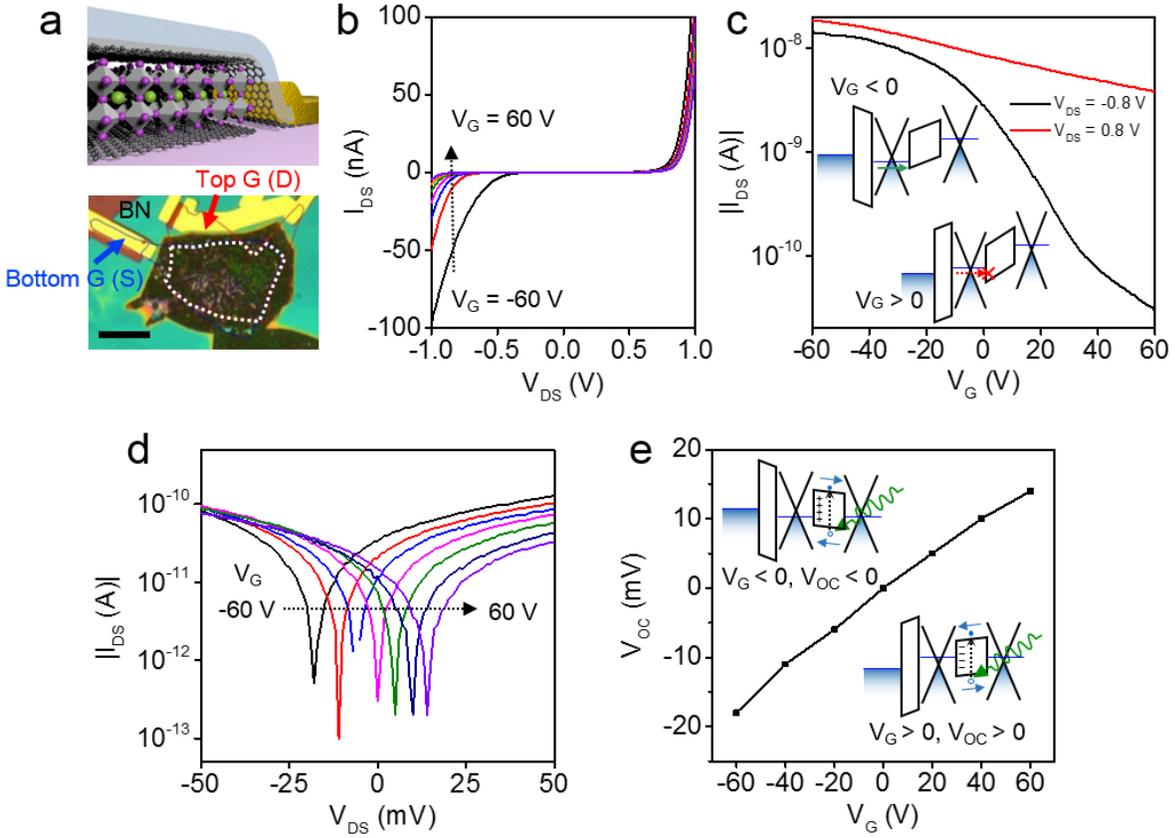

**Figure. 3.** Optoelectronic characterization on BN-covered GPG device at 77K. (a) The device schematic and optical image of a BN-covered GPG device. The BN-covered top- and bottom-graphene were connected to the metal electrodes by edge contacts. The red and blue lines framed the top- and bottom-graphene, respectively; and the white dashed line outlines the GPG overlapping area. The bias is applied to top-graphene while the bottom graphene is grounded during all measurement. The black scale bar is 10 μm. (b) The dark $I_{DS}$-$V_{DS}$ characteristics measured under different gate voltage. The negatively-biased regime show much bigger gate modulation than the positively-biased regime. (c) The transfer characteristics of the device when biased at -0.8 V (black) and 0.8 V (red). The negatively-biased curve shows a larger on/off ratio of ~500 compared to positively-biased one which is only about ~5. The two band diagrams show negatively-biased condition under positive and negative gate voltage. As $V_G < 0$ (> 0), the hole-injection barrier decreases (increases), the device current increases (decreases). The green arrow indicates an effective hole-injection while the red dashed arrow indicates the blocked hole injection. (d) The logarithmic plot of $I_{DS}$-$V_{DS}$ curves obtained by sweeping at smaller bias voltage range under continuous laser illumination. The open-circuit voltage ($V_{OC}$) was obtained from current minima and able to be modulated from negative to positive while the gate voltage is varied from -60 V to 60 V. The dashed arrow indicates the order of gate-voltage varying sequence. (e) The gate-dependent $V_{OC}$ extracted from (d). The band diagrams depict the gate-tunable $V_{OC}$ and the ion accumulation effect that reduces the gate-modulation of photovoltaic effect in GPG device.

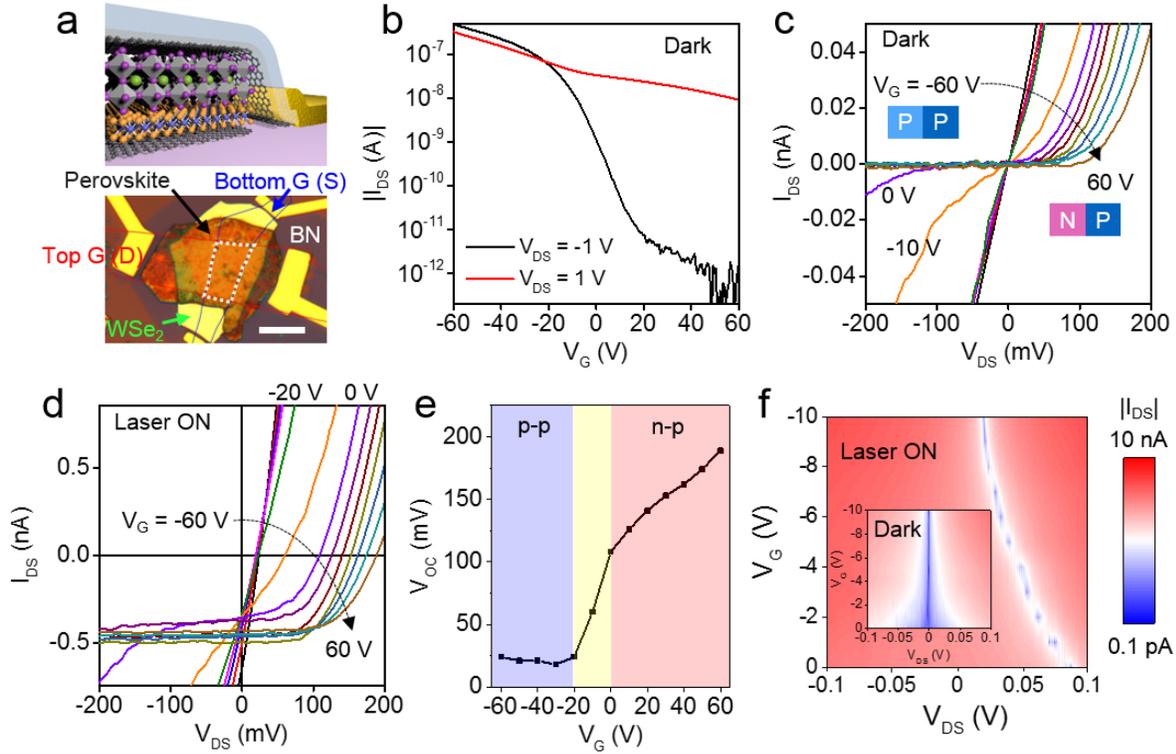

**Figure 4.** Optoelectronic characteristics of graphene/WSe$_2$/perovskite/graphene (GWPG) heterostructure devices. (a) The schematic illustration and optical image of a BN-covered GWPG device. The red and blue lines framed the top- and bottom-graphene, respectively; the perovskite and WSe$_2$ are indicated by black and green arrows, respectively; the white dashed lines outline the overlapping area of GWPG structure. The top-graphene is biased while the bottom graphene is grounded for all measurements. (b) The transfer characteristics of the GWPG device when biased at -1 V (black) and 1 V (red). The negatively-biased curve shows a large on/off ratio around $10^6$ while the positively-biased one is only ~35. (c) The dark $I_{DS}$-$V_{DS}$ characteristics of GWPG device show an obvious transition from symmetric p-p junction to rectifying n-p diode as $V_G$ varied from -60 V to 60 V (each curve is 10 V increment). In the negative bias range, the $I_{DS}$ current shows a dramatic decrease from $V_G$ = -10 V (orange) to 0 V (purple). (d) The $I_{DS}$-$V_{DS}$ obtained under laser 532 nm illumination with varied $V_G$ (10 V increment). (e) The gate-dependent $V_{OC}$ extracted from (d). The blue area is p-p junction regime ($V_G$ = -60 V to -20 V) where $V_{OC}$ is small with little gate-modulation; the yellow area ($V_G$ = -20 V to 0 V) is the transition region where $V_{OC}$ shows dramatic change with gate voltage; the red area is n-p junction region where $V_{OC}$ increases slightly less than transition area with gate voltage. (f) The logarithmic plot of laser-illuminated $I_{DS}$-$V_{DS}$ curves under different gate voltage within transition region. A significant open-circuit voltage $V_{OC}$ (current minima) shift was observed as the gate bias varied, while the dark $I_{DS}$-$V_{DS}$ plots shown in the inset has no change of current minima with the gate bias. All measurements are done at 77K.